\documentclass[]{aa}  

\usepackage{graphicx}
\usepackage{txfonts}
\usepackage{natbib}
\usepackage{bbm}
\usepackage[breaklinks=true]{hyperref}
\bibpunct{(}{)}{;}{a}{}{,}             
\begin{document}

   \title{Galaxy filaments as pearl necklaces}


   \author{E.~Tempel\inst{1,2}
		  \and R.~Kipper\inst{1,3}
		  \and E.~Saar\inst{1,4}
		  \and M.~Bussov\inst{1,5}
		  \and A.~Hektor\inst{2}
		  \and J.~Pelt\inst{1}
          }

   \institute{Tartu Observatory, Observatooriumi~1, 61602 T\~oravere, Estonia\\
             \email{elmo.tempel@to.ee}
			 \and National Institute of Chemical Physics and Biophysics, R\"avala pst 10, 10143 Tallinn, Estonia
			 \and Institute of Physics, University of Tartu, 51010 Tartu, Estonia
			 \and Estonian Academy of Sciences, Kohtu 6, 10130 Tallinn, Estonia
			 \and Institute of Mathematical Statistics, University of Tartu, 50409 Tartu, Estonia
             }


 
  \abstract
   {Galaxies in the Universe form chains (filaments) that connect groups and clusters of galaxies. The filamentary network includes nearly half of the galaxies and is visually the most striking feature in cosmological maps.}
   {We study the distribution of galaxies along the filamentary network, trying to find specific patterns and regularities.}
   {Galaxy filaments are defined by the Bisous model, a marked point process with interactions.  We use the two-point correlation function and the Rayleigh $Z$-squared statistic to study how galaxies and galaxy groups are distributed along the filaments.}
   {We show that galaxies and groups are not uniformly distributed along filaments, but tend to form a regular pattern. The characteristic length of the pattern is around 7~$h^{-1}$Mpc. A slightly smaller characteristic length 4~$h^{-1}$Mpc can also be found, using the $Z$-squared statistic.}
   {We find that galaxy filaments in the Universe are like pearl necklaces, where the pearls are galaxy groups distributed more or less regularly along the filaments. We propose that this well defined characteristic scale could be used to test various cosmological models and to probe environmental effects on the formation and evolution of galaxies.}


   \maketitle
%
\section{Introduction}

	Many cosmological probes have been developed and used to estimate the parameters of the cosmological models describing our Universe. Most of them rely on some aspects of the large-scale structure. Most common probe to quantify galaxy clustering is the two-point correlation function which has been used already decades ago \citep{Davis:76, Groth:77, Davis:88, Hamilton:88, White:88, Boerner:89, Einasto:91}. Some recent examples include \citet{Connolly:02}, \citet{Zehavi:05}, \citet{Contreras:13}, \citet{deSimoni:13}, \citet{Wang:13}, and references in these papers. Other examples of cosmological probes are the three-point correlation functions \citep[e.g.][]{McBride:11, Marin:13, Guo:14} and the power spectrum analysis \citep{Hutsi:06, Hutsi1:10, Blake:10, Balaguera:11}.

   	It is well known that galaxy filaments are visually the most dominant structures in the galaxy distribution, being part of the so-called cosmic network \citep{Joeveer:78,Bond:96}. Presumably, nearly half ($\sim\!40\%$) of the galaxies (or mass in simulations) are located in filaments. This number is based on morphological or dynamical classification of the cosmic web \citep[e.g.][]{Jasche:10,Tempel:14a,Cautun:14}. Properties of the three-point correlation function also indicate that galaxies tend to populate filamentary structures \citep{Guo:14} and, indeed, filaments have been found between galaxy clusters \citep[e.g.][]{Dietrich:12}. \citet{Tempel:14b} shows that filaments extracted from the spatial distribution of galaxies/haloes are also dynamical structures that are well connected with the underlying velocity field. Galaxy filaments that connect groups and clusters of galaxies are also affecting the evolution of galaxies \citep[e.g.][]{Tempel:13a,Tempel:13b,Zhang:13} and the distribution of satellites around galaxies \citep{Guo:14a}.
	
	In this paper we study the distribution of galaxies along galaxy filaments to search for regularities in galaxy and group distributions. Such a clustering pattern exists at least in some filaments as shown decades ago \citep[e.g.][]{Joeveer:78,Einasto:80}. In the current study, we use the two-point correlation function \citep[see][]{Peebles:80} and the Rayleigh $Z$-squared statistics \citep[see][]{Buccheri:92} to look for the regularity in the galaxy distribution. We suggest that the regularities we find in the galaxy distribution could be used as cosmological probes for dark energy and dark matter, the mysterious components in the dark energy dominated cold dark matter ($\Lambda$CDM) cosmological models.
   
   	Throughout this paper we assume the Wilkinson Microwave Anisotropy Probe (WMAP) cosmology: the Hubble constant $H_0 = 100\,h\ \mathrm{km\,s^{-1}Mpc^{-1}}$, with $h=0.697$, the matter density $\Omega_\mathrm{m}=0.27$ and the dark energy density $\Omega_\Lambda=0.73$ \citep{Komatsu:11}.

\section{Data and methods}

\subsection{Galaxy and group samples for the SDSS data}

The present work is based on the Sloan Digital Sky Survey \citep[SDSS,][]{York:00} data release~10 \citep[DR10;][]{Ahn:13}. We use the galaxy and group samples as compiled in \citet{Tempel:14} that cover the main contiguous area of the survey (the Legacy Survey, approximately 17.5\% from the full sky). The flux-limited catalogue extends to the redshift 0.2 (574~$h^{-1}$Mpc) and includes 588193 galaxies and 82458 groups with two or more members. 

In \citet{Tempel:14} the redshift-space distortions, the so-called finger-of-god (FoG) effects, are suppressed using the rms sizes of galaxy groups in the plane of the sky and their rms radial velocities as described in \citet{Liivamagi:12}. We calculate the new radial distances for galaxies in groups in order to make the galaxy distribution in groups approximately spherical. We note that such a compression will remove the artificial line-of-sight filament-like structures as shown in \citet{Tempel:12}.

\subsection{Bisous model: extracting filaments from galaxy distribution}

The detection of filaments is performed by applying an object/marked point process with interactions \citep[the Bisous process;][]{Stoica:05} to the distribution of galaxies. This algorithm provides a quantitative classification which agrees with the visual impression of the cosmic web and is based on a robust and well-defined mathematical scheme. A detailed description of the Bisous model is given in \citet{Stoica:07,Stoica:10} and \citet{Tempel:14a}. For reader convenience, a brief and intuitive description is given below.

The marked point process we propose for filament detection is different from the ones already used in cosmology. In fact, we do not model galaxies, but the structure outlined by galaxy positions.

This model approximates the filamentary network by a random configuration of small segments (thin cylinders). We assume that locally galaxies may be grouped together inside a rather small cylinder, and such cylinders may combine to form a filament if neighbouring cylinders are aligned in similar directions. This approach has the advantage that it is using only positions of galaxies and does not require any additional smoothing to create a continuous density field.

The solution provided by our model is stochastic. Therefore, we find some variation in the detected patterns for different Markov chain Monte Carlo (MCMC) runs of the model. The main advantage of using such a stochastic approach is the ability to give simultaneous morphological and statistical characterisation of the filamentary pattern.

In practice, after fixing the approximate scale of the filaments, the algorithm returns the filament detection probability field together with the filament orientation field. Based on these data,  filament spines are extracted and a filament catalogue is built. Every filament in this catalogue is represented as a spine: a set of points that define the axis of the filament.

The spine detection we use is based on two ideas. First, filament spines are located at the highest density regions outlined by the filament probability maps. Second, in these regions of high probability for the filamentary network, the spines are oriented along the orientation field of the filamentary network. See \citet{Tempel:14a} for more details of the procedure.

\begin{figure}
   \centering
   \includegraphics[width=80mm]{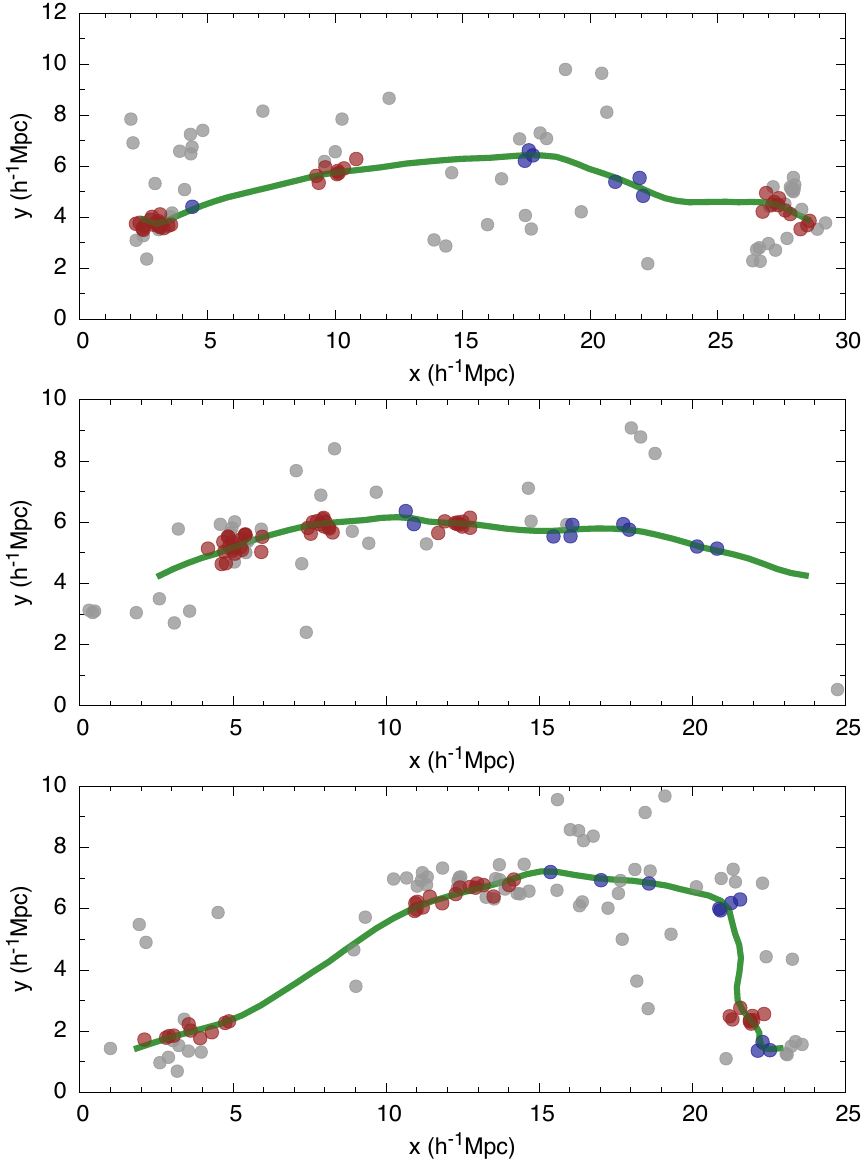}
   \caption{Examples of filaments and their spines. Red points show galaxies in filaments (closer than 0.5~$h^{-1}$Mpc to the filament axis) that are located in groups with 5 or more members. The group richness is taken from \citet{Tempel:14}. Blue points show other galaxies in filaments. Grey points are background galaxies that are not located in these filaments. Note that some of the background galaxies project to the filament spines, but are actually located farther than 0.5~$h^{-1}$Mpc from the spine. The thick green line shows the spine of a filament. To show the scale of structures, cartesian coordinates are shown around each filament.}
   \label{fig:fil}
\end{figure}

\subsection{Galaxy filament sample for the SDSS data}

The catalogue of filaments is built by applying the Bisous process to the distribution of galaxies as outlined above. The method and parameters are exactly the same as in \citet{Tempel:14a}, where the Bisous model was applied to the SDSS DR8 \citep{Aihara:11} data. Since the datasets in the SDSS DR8 and DR10 are basically the same, the extracted filaments in DR10 are statistically the same as presented in \citet{Tempel:14a}. The assumed scale (radius) for the extracted filaments is 0.5~$h^{-1}$Mpc. Because the survey is flux-limited, the sample is very diluted farther away. Hence, we are only able to detect filaments in this scale up  to the redshift 0.15 ($\approx 450$~$h^{-1}$Mpc). The longest filaments in our sample are up to 70~$h^{-1}$Mpc long.

\begin{figure}
   \centering
   \includegraphics[width=83mm]{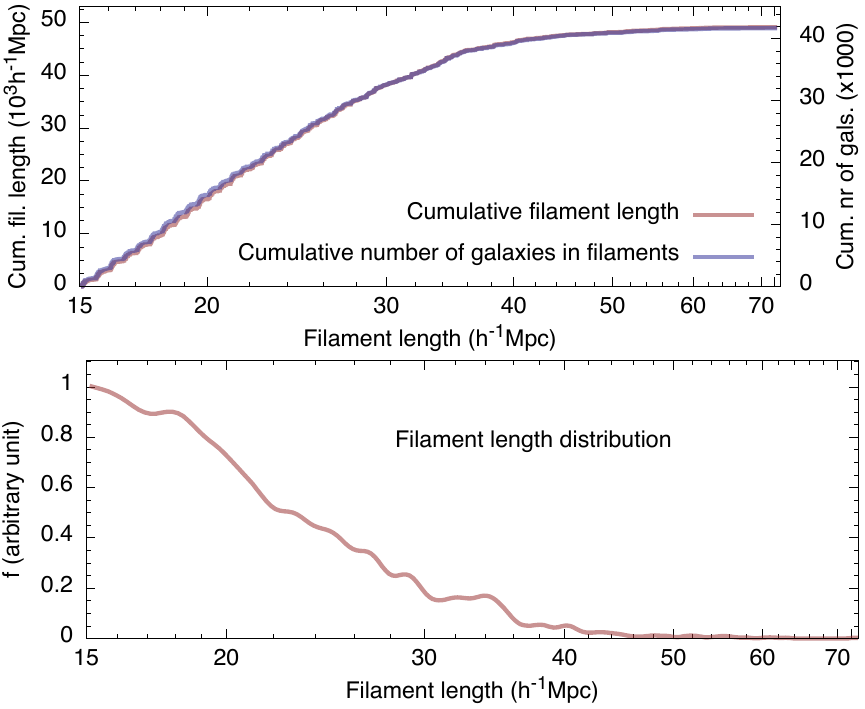}
   \caption{\emph{Upper panel:} the cumulative filament length and the cumulative number of galaxies in filaments as a function of filament length: the lines show the sum of the filament lengths (red line) and the number of galaxies in filaments (blue line) summed over filaments shorter than indicated in the abscissa. On average there is slightly less than one galaxy per $h^{-1}$Mpc. \emph{Lower panel:} the filament length distribution in our sample.}
   \label{fig:filstat}
\end{figure}

The filaments are extracted from a flux-limited galaxy sample, hence, the completeness of extracted filaments decreases with distance. In \citet{Tempel:14a} we showed that the volume filling fraction of filaments is roughly constant with distance if filaments longer than 15~$h^{-1}$Mpc are considered.  Therefore, in the current study we are using only filaments longer than this limit.  In addition, this choice is justified, since longer filaments allow us to study the distribution of galaxies along the filaments, which is the purpose of this paper. The remaining incompleteness of filaments in our sample is not a problem, because we are analysing single filaments. The filaments we use are the strongest filaments (or segments of filaments) in the sample.

To suppress the flux-limited sample effect, we volume-limit the galaxy content for single filaments. For that, we find for every filament the maximum distance (from the observer) of its galaxies and the corresponding magnitude limit and use only galaxies brighter than that. Since the majority of filaments extend over a relatively narrow distance interval (always narrower than the length of the filament), the number of excluded galaxies in every filament is small.

To study the galaxy spacing along the filaments, we project every galaxy to the filament spine. Hence, the distance is measured along the filament spine. Figure~\ref{fig:fil} illustrates extracted galaxy filaments and their spines in the field of galaxies.

For our analysis, we use only filaments that contain at least 10 galaxies. When studying galaxy groups, we use only filaments where the number of groups per filament is at least 5. We use the groups from the catalogue by \citet{Tempel:14}. The upper panel in Fig.~\ref{fig:filstat} shows the cumulative filament length and the cumulative number of galaxies in filaments as a function of the filament length.

The lower panel in Fig.~\ref{fig:filstat} shows the filament length distribution. In this study, a galaxy is considered to belong to a filament if it is closer than 0.5~$h^{-1}$Mpc to the filament spine. This is also the scale of detected filaments in our Bisous model. In addition, we study the distribution of galaxies that are closer than 0.25~$h^{-1}$Mpc to the filament spine. In Fig.~\ref{fig:filstat} we see that the total filament length in our study is 50000~$h^{-1}$Mpc and half of it comes from the filaments shorter than 23~$h^{-1}$Mpc. The numbers of filaments and galaxies in filaments for studied subsamples are given in Table~\ref{tab:stat}.

\begin{table}
\caption{The numbers of filaments ($N_\mathrm{fil}$) and galaxies in filaments ($N_\mathrm{gal}$) in various subsamples. Only filaments longer than 15~$h^{-1}$Mpc and containing at least 10 galaxies (or 5 groups) are considered. $N_\mathrm{rich}$ denotes the number of galaxies in groups (richness) and $d_\mathrm{fil}$ is the galaxy/group distance from the filament spine.}
\label{tab:stat}
\centering
\begin{tabular}{lcc}
\hline\hline
Sample & $N_\mathrm{fil}$ & $N_\mathrm{gal}$ \\
\hline
All galaxies ($d_\mathrm{fil}<0.5~h^{-1}$Mpc) & 2150 & 41543 \\
All galaxies ($d_\mathrm{fil}<0.25~h^{-1}$Mpc) & 1752 & 26830 \\
Groups ($N_\mathrm{rich}\geq 1$; $d_\mathrm{fil}<0.5~h^{-1}$Mpc) & 1943 & 15874 \\
Groups ($N_\mathrm{rich}\geq 1$; $d_\mathrm{fil}<0.25~h^{-1}$Mpc) & 1493 & 10492 \\
Groups ($N_\mathrm{rich}\geq 2$; $d_\mathrm{fil}<0.5~h^{-1}$Mpc) & 759 & 4667 \\
Groups ($N_\mathrm{rich}\geq 2$; $d_\mathrm{fil}<0.25~h^{-1}$Mpc) & 401 & 2323 \\
Groups ($N_\mathrm{rich}\geq 3$; $d_\mathrm{fil}<0.5~h^{-1}$Mpc) & 211 & 1196 \\
Groups ($N_\mathrm{rich}\geq 3$; $d_\mathrm{fil}<0.25~h^{-1}$Mpc) & 111 & 606 \\
\hline
\end{tabular}
\end{table}

\subsection{Estimating the pair correlation function}

To study galaxy correlations in filaments, we use the two-point correlation function $\xi(\vec{r})$ that measures the excess probability of finding two points separated by a vector $\vec{r}$ compared to that probability in a homogeneous Poisson sample \citep{Peebles:80, Martinez:02}. For galaxy filaments, the correlation function along the filament can be expressed in terms of the distances between the galaxies measured along the filament. Note that this is not exactly the case when studying a sample from a galaxy redshift survey. The line-of-sight component of the position of a galaxy is derived from the observed redshift, hence, the distance along the line of sight is influenced by redshift distortions. However, since we are interested in scales larger than the group/cluster scale, we can ignore this effect. We measure the distances between galaxies along filament spines. Below (see Fig.~\ref{fig:cor_all}) we divide our filament sample into parallel to the line of sight and perpendicular to the line of sight subsamples and show that redshift space distortions do not affect our results.

We estimate $\xi(r)$ following the Landy-Szalay border-corrected estimator \citep{Landy:93}. We generated a random distribution of points for each filament considered, and estimated the correlation function $\xi(r)$:
\begin{equation}
	\widehat{\xi}(r) = 1+\frac{DD(r)}{RR(r)}-2\frac{DR(r)}{RR(r)},
	\label{eq:xi1}
\end{equation}
where $DD(r)$, $RR(r)$, and $DR(r)$ are the probability densities for the galaxy-galaxy, random-random, and galaxy-random pairs, respectively, for a pair distance $r$. The random catalogue is built separately for each filament assuming an uniform distribution of points along the spine of the filament. In this work, we fix the size of the random point set for each filament by $N_\mathrm{rd}=50N$ ($N$ is the number of galaxies in each filament). We tested that our results do not change if we increase the number of random points used to  $N_\mathrm{rd}=100N$.

We estimate the probability densities by the kernel method, summing the box spline $B_3(\cdot)$ kernels \citep{Saar:07} centred at each pair distance, and sampling the distributions at smaller intervals than the kernel width. The kernel width we use is $1.0$~$h^{-1}$Mpc (for groups it is twice as large). In \citet{Martinez:09} the same method is applied to detect baryon acoustic peaks. Such a method does not require binning the data and with a good choice of the kernel width, can optimally recover the probability distribution; it does not introduce any bias in the correlation estimator. The kernel width influences the estimate the same way as the bin width -- if the kernel is too wide, the signal is lost, and in the contrary, if the kernel is too narrow, noise dominates. See Appendix~\ref{app:kernel} for the details of the kernel method and for comparison with binning the data.

Clustering in configuration space is subject to the integral constraint which introduces a bias (i.e., a shift in the overall shape) in the estimates of the correlation function. Since the lengths of the filaments are comparable with the studied scale of the correlation function and the length of each filament is different (hence the bias is different), we cannot ignore the integral constraint. The first-order approximation for the integral constraint can be estimated as explained in \citet{Roche:99} and \citet{Labatie:12}.

\begin{figure}
   \centering
   \includegraphics[width=80mm]{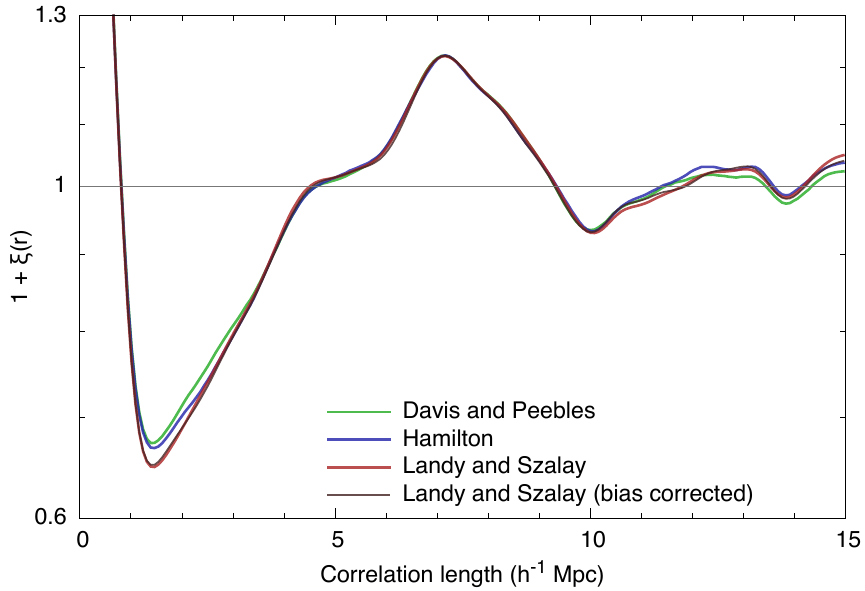}
   \caption{Galaxy correlation function along filaments. Various correlation function estimates are shown: \citet{Davis:83} (green line), \citet{Hamilton:93} (blue line), and \citet{Landy:93} (red line). The black line shows the integral constraint corrected correlation function computed using the Landy-Szalay estimator.}
   \label{fig:cor_test}
\end{figure}

In all cases studied below, we found that the integral constraint correction does not change the features of our measured correlation function, it mostly affects only the general scaling of the correlation function. Figure~\ref{fig:cor_test} shows the Landy-Szalay correlation function estimator without (red) and with (black) integral constraint correction. The differences are irrelevant, hence, our results are not sensitive to the details of the estimation of the integral constraint.

Additionally, we checked how the estimated correlation function depends on the used estimator. In Fig.~\ref{fig:cor_test} we show the correlation function estimated using the \citet{Davis:83} and \citet{Hamilton:93} estimators. The differences between the estimates are very small, and they do not affect the features in the correlation function that we are studying.

Using Eq.~(\ref{eq:xi1}), we estimate the correlation function for each filament separately. The final pair correlation function, averaged over all filaments, is estimated as a weighted sum
\begin{equation}
	\xi^\mathrm{fil}(r) = \frac{\sum\limits_{i=1}^{N_\mathrm{fil}} \mathbbm{1}\{L_{\mathrm{fil},i}>r\}L_{\mathrm{fil},i} \, \xi_i(r)} {\sum\limits_{i=1}^{N_\mathrm{fil}} \mathbbm{1}\{L_{\mathrm{fil},i}>r\}L_{\mathrm{fil},i}},
\end{equation}
where summation is over all filaments, $\xi_i$ is the correlation function for a single filament, $L_{\mathrm{fil},i}$ is the length of the $i$th filament, and $\mathbbm{1}\{\cdot\}$ is the indicator function that selects only filaments longer than $r$.

We estimate the statistical error on our $\xi^\mathrm{fil}(r)$ measurements with the standard jackknife method \citep[see e.g.][]{Norberg:09}, where we omit one filament under consideration at a time. In our figures we show the 95\% confidence intervals.

\subsection{Rayleigh ($Z$-squared) statistic}

To test whether the galaxy distribution might have regularity along the filaments we use the Rayleigh (or $Z$-squared) statistic \citep[see e.g.][]{Buccheri:83, Muno:03}. It is an excellent method when the event rate (in our case, the number of galaxies per filament) is low. The method has been used to detect periodicity in time series for the data in the form of discrete events (photon arrival times) and can be applied to detect periodicity in the galaxy distribution along filaments, where the galaxy positions can be considered as events.

The algorithm works as follows. For each filament, we produce a periodogram using the $Z^2_1$ (Rayleigh statistic), 
\begin{equation}
	Z_1^2 = \frac{2}{N}\left[\left(\sum\limits_{j=1}^N \cos\phi_j\right)^2 + \left(\sum\limits_{j=1}^N \sin\phi_j\right)^2\right] ,
\end{equation}
where $N$ is the number of galaxies in a filament and $\phi_j = 2\pi l_j/d$ is the phase value for a galaxy $j$ for a fixed period $d$; $l_j$ is a distance of the galaxy $j$ along the filament spine from the beginning of the filament.

To measure the $Z$-squared statistic for a period $d$, we are using only filaments longer than $2d$. This assures that there are at least two periods for each filament. For a signal resulting purely from a Poisson noise, $Z^2_1$ has a $\chi^2$ distribution with two degrees of freedom. However, this is only in the case if the number of events is high enough. If the number of events is lower than 100 (as usual in our case), $Z^2_1$ does not have a $\chi^2$ probability distribution. So, we are deriving a null-hypothesis probability function using Monte Carlo simulations with $N$ (the number of galaxies in a filament) data points assuming an uniform distribution of points along the spine of the filament. This allows us also to estimate the confidence intervals for our measured signal.

We compute the $Z^2_1$ statistic as a function of a period $d$ for every filament and then we find the average signal using the filament length $L_\mathrm{fil}$ as the weight
\begin{equation}
	Z_{1,\mathrm{fil}}^2(d) = \frac{\sum\limits_{i=1}^{N_\mathrm{fil}} \mathbbm{1}\{L_{\mathrm{fil},i}>2d\}L_{\mathrm{fil},i} \, Z_{1,i}^2(d)} {\sum\limits_{i=1}^{N_\mathrm{fil}} \mathbbm{1}\{L_{\mathrm{fil},i}>2d\}L_{\mathrm{fil},i}} ,
\end{equation}
where $Z_{1,i}^2$ is the $Z$-squared statistic for the $i$-th filament. The confidence limits are found using a jackknife technique, as we did for the correlation functions.

Examples of how the $Z$-squared statistic works are given in Appendix~\ref{app:z2}.

\section{Results}

\subsection{Correlations along galaxy filaments}

\begin{figure}
   \centering
   \includegraphics[width=88mm]{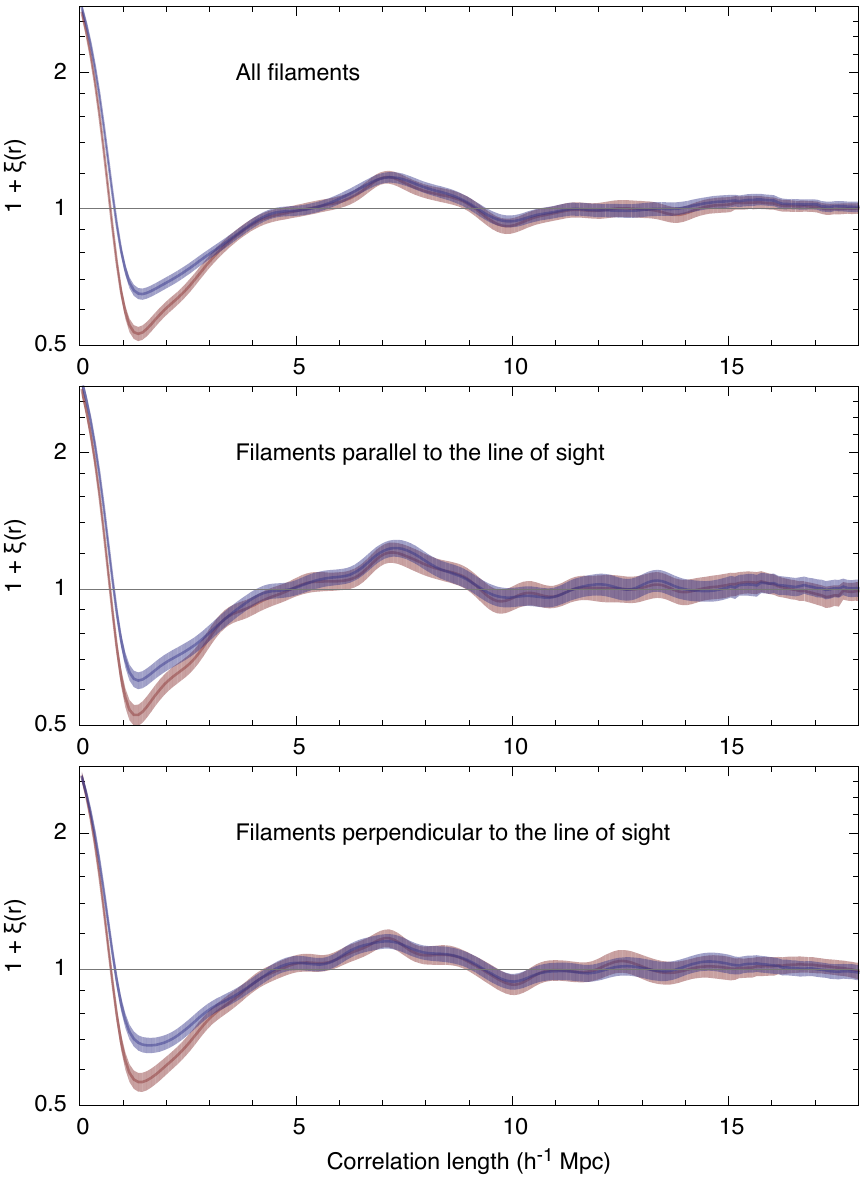}
   \caption{Galaxy correlation function along filaments. The upper panel shows the correlation function for all filaments, the middle panel shows this function for filaments that are parallel to the line-of-sight ($\cos i >0.5$), and the lower panel shows the correlation function for filaments that are perpendicular to the line-of-sight ($\cos i < 0.5$): $i$ denotes the angle between the line-of-sight and the filament direction. Blue and red lines with filled regions show the correlation function together with its 95\% confidence limits (based on a jackknife estimate) for galaxies closer than 0.5 and 0.25~$h^{-1}$Mpc to the filament axis, respectively.}
   \label{fig:cor_all}
\end{figure}

\begin{figure}
   \centering
   \includegraphics[width=88mm]{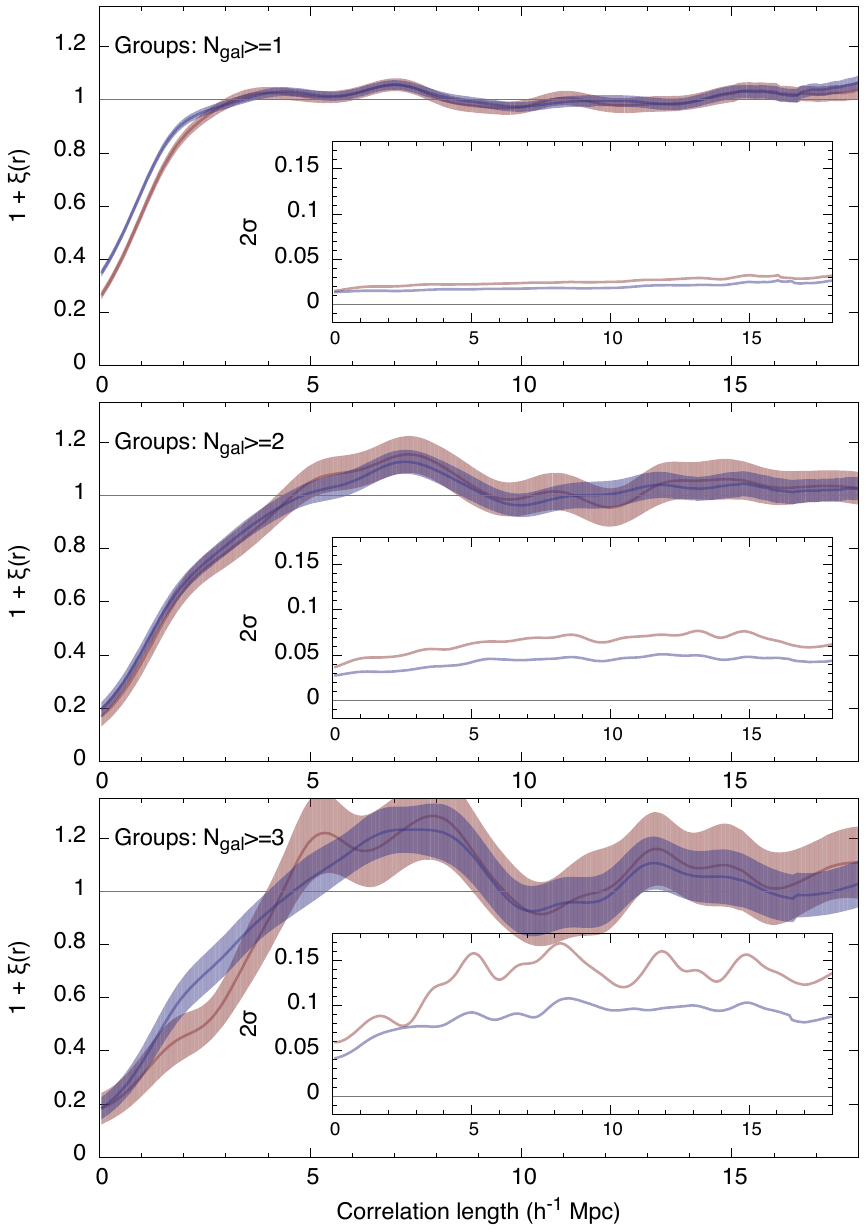}
   \caption{Galaxy correlation function along filaments for groups. The upper panel shows the correlation function for all groups and isolated galaxies together, the middle panel shows the correlation function  only for groups with two or more members, and the bottom panel for groups with at least three member galaxies. The group richness (number of galaxies in a group) is taken from \citet{Tempel:14}. Blue lines show the correlation function for groups closer than 0.5~$h^{-1}$Mpc to the filament axis and red lines show the correlation function for groups closer than 0.25~$h^{-1}$Mpc. The shaded regions are the 95\% confidence intervals. In the inset panels, the width of the 95\% confidence regions as a function of the pairwise  distance is shown.}
   \label{fig:cor_groups}
\end{figure}

Figure~\ref{fig:cor_all} shows the galaxy correlation function along filaments. Three specific features are seen in this correlation function: a maximum near the zero pair distance, a minimum that follows it, and a bump close to 7~$h^{-1}$Mpc. The first maximum is caused by galaxy groups. It shows that galaxies are not distributed uniformly in the space, they form groups and clusters, as it is well known. The minimum next to the first maximum shows that groups themselves are not distributed uniformly along filaments. It shows that two groups cannot be located directly close to each other (merging groups are exceptions) and there exists some preferred minimum distance between galaxy groups. This is also expected since matter is falling into groups and there is not enough matter in a close-by neighbourhood of groups to form another group. The most interesting feature is the small maximum close to 7~$h^{-1}$Mpc. It shows that galaxies (and also groups) have a tendency to be separated by this distance along the filaments. We will analyse the nature of this bump in more detail below.

Since we are using a galaxy sample where FoG effects are suppressed, we checked whether and how this influences our main results. For that we divided our filament sample into two subsamples: mostly parallel to the line-of-sight ($\cos i >0.5$) and mostly perpendicular to the line-of-sight ($\cos i <0.5$). Here $i$ denotes the angle between the direction of the filament and the line-of-sight. The correlation functions for these two subsamples are shown in the middle and lower panels in Fig.~\ref{fig:cor_all}, respectively. We note that the bump around 7~$h^{-1}$Mpc is visible in both subsamples, hence, it is a real feature. However, the bump is slightly weaker for the filaments perpendicular to the line of sight. So, we also calculated the correlation function for filaments where $\cos i <0.25$, and the bump remained in the same place. We conclude that the 7~$h^{-1}$Mpc maximum is not affected by the FoG effect. The most noticeable difference between the middle and lower panels in Fig.~\ref{fig:cor_all} is that the minimum is slightly stronger for filaments parallel to the line-of-sight. This is probably because of our FoG suppression. When suppressing FoG effects, we also suppress the line-of-sight scatter of some of the background and foreground galaxies in the group neighbourhood that creates a lack of galaxies close to the groups. To test the effect of FoG suppression, we calculated the correlation signal for galaxies closer than 0.5 and 0.25~$h^{-1}$Mpc to the filament axis (blue and red lines in Fig.~\ref{fig:cor_all}, respectively). For the smaller radius, the effect of finger-of-god suppression should be smaller. As expected, the prominent features are visible in both subsamples.

In Fig.~\ref{fig:cor_groups} we show the correlation function for galaxy groups. In this figure, the smoothing scale is twice as large as used for the  galaxy correlation functions and we analyse only filaments that contain at least five groups. Figure~\ref{fig:cor_groups} shows the correlation function for groups with different minimum group richness. We see that the bump around 7~$h^{-1}$Mpc is present in every subsample, indicating that groups themselves have some preferred distance between them. Interestingly, the upper panel in Fig.~\ref{fig:cor_groups} shows a small bump around 4~$h^{-1}$Mpc as well. This scale is also seen when using the $Z$-squared statistic (see Sect.~\ref{sect:res_ray}).

The biggest difference when comparing the correlation function for galaxies (Fig.~\ref{fig:cor_all}) and for groups (Fig.~\ref{fig:cor_groups}) is the fact that there is no zero-distance maximum for groups. This shows that groups themselves do not form clusters, which is expected. The minimum close to zero distance in the group correlation function shows that there is a lack of other groups around each group. The radius of the group influence extends to 4~$h^{-1}$Mpc, after that the correlation function is mostly flat. 

\begin{figure}
   \centering
   \includegraphics[width=88mm]{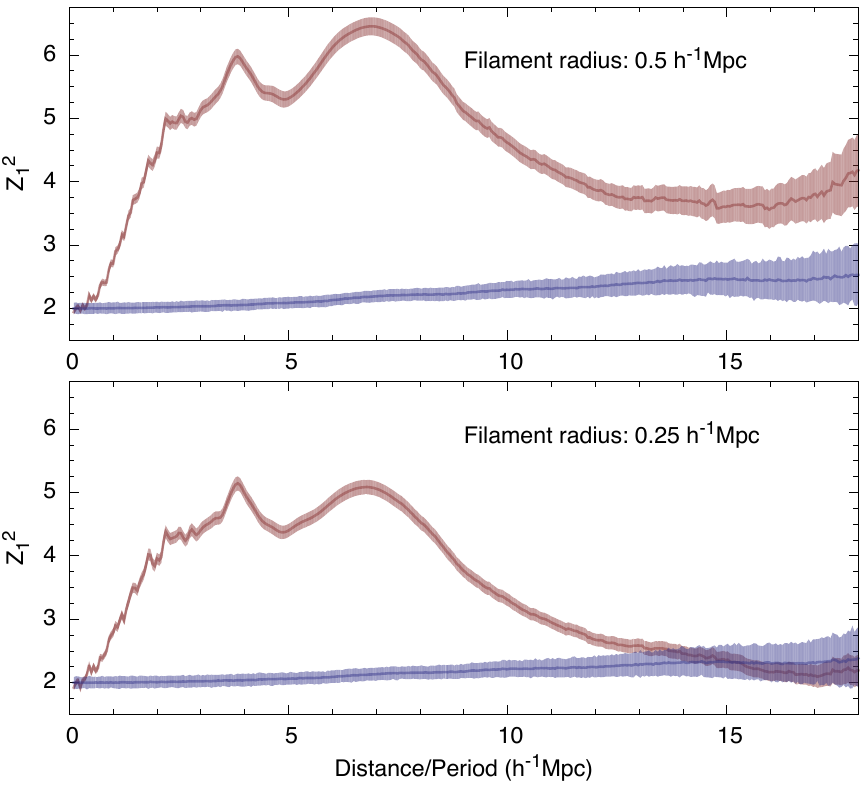}
   \caption{The dependence of the Rayleigh ($Z$-squared) statistic $Z^2_1$ on distance (period). The upper panel shows the results for galaxies closer than 0.5~$h^{-1}$Mpc to the filament axis and the lower panel shows the results for galaxies closer than 0.25~$h^{-1}$Mpc. The red line shows the $Z^2_1$ statistic together with the jackknife 95\% confidence estimate. The blue line shows the results from Monte Carlo simulations for the null hypothesis together with the 95\% confidence limits.}
   \label{fig:ray_all}
\end{figure}

Figure~\ref{fig:cor_groups} also shows that the strength of the bump at 7~$h^{-1}$Mpc is higher if isolated galaxies ($N_\mathrm{gal}=1$) are excluded. This indicates that the bump is related to galaxy clustering. Unfortunately, the current dataset is too small to adequately quantify the dependence of the amplitude and significance of the bump on group richness.

Additionally, we checked how the correlation signal depends on the galaxy luminosity, colour or morphology. We found a weak dependency on colour and luminosity, so that the bump at 7~$h^{-1}$Mpc is slightly stronger for redder and more luminous galaxies. The interpretation of this is unclear. On one hand, it could be related to galaxy evolution and in this case could provide information on galaxy bias. On the other hand, it might be just the result of galaxy clustering. More luminous and redder galaxies are more clustered and the bump at 7~$h^{-1}$Mpc is also stronger for groups.

\subsection{Regularity of the galaxy distribution along filaments}
\label{sect:res_ray}

\begin{figure}
   \centering
   \includegraphics[width=88mm]{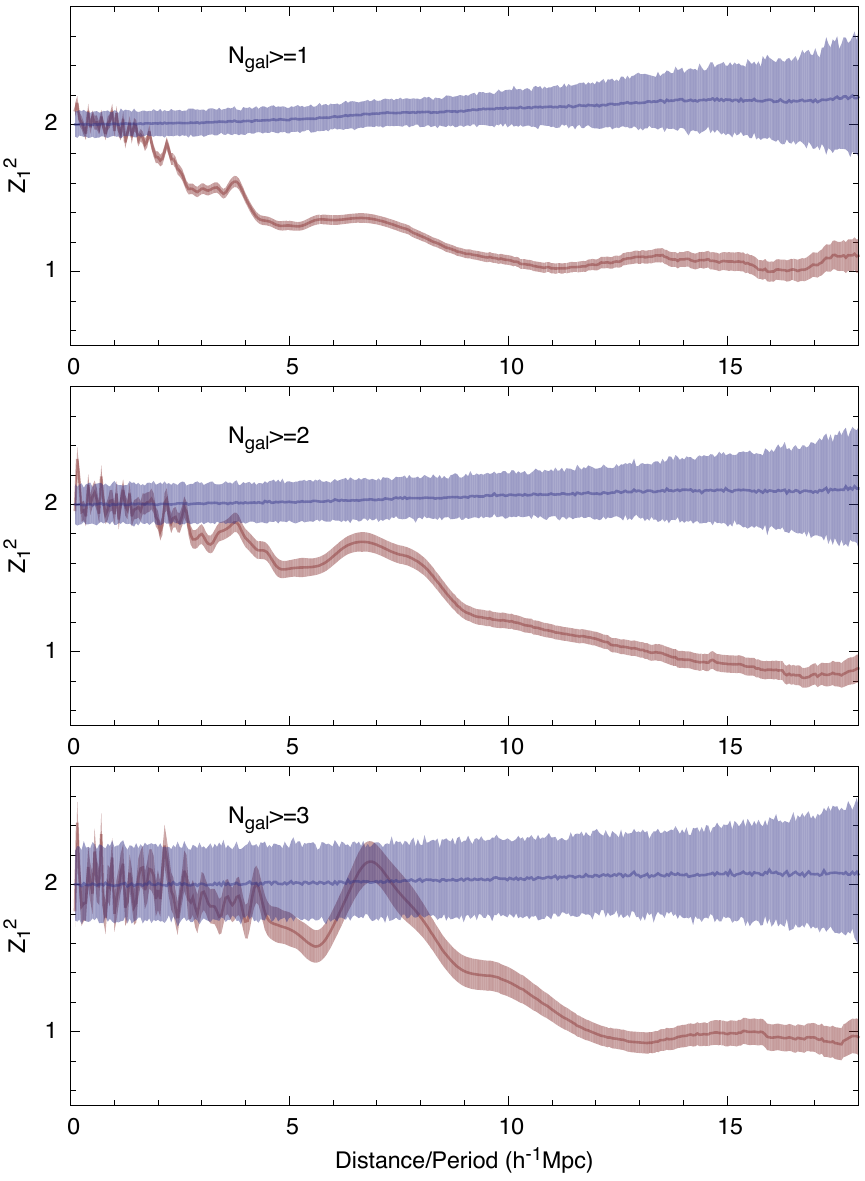}
   \caption{The $Z$-squared statistic for groups, where the minimum group richness is one (upper panel), two (middle panel), or three (lower panel), respectively. The designation of lines is the same as in Fig.~\ref{fig:ray_all}.}
   \label{fig:ray_cl}
\end{figure}

Figure~\ref{fig:ray_all} shows the $Z$-squared statistic for all galaxies closer than 0.5~$h^{-1}$Mpc (upper panel) and 0.25~$h^{-1}$Mpc (lower panel) to the filament axis. The $Z$-squared statistic based on galaxies is shown with red lines, where the shaded region shows the 95\% confidence limits. The blue line shows the statistic for the null hypothesis using Monte Carlo simulation for a Poisson sample. The shaded region shows the 95\% confidence limits for this case. For Monte Carlo simulation, the filaments and numbers of galaxies per filament are the same as for the real sample, but galaxies are Poisson distributed. Since we are averaging filaments with different lengths, the  $Z$-squared statistic for the Poisson sample is not exactly around 2 as in Fig.~\ref{fig:ray_test}. Since the deviation is small, it does not affect our conclusions.

The maximum around 7~$h^{-1}$Mpc that was visible in the correlation functions is also visible in Fig.~\ref{fig:ray_all}, confirming that galaxies are distributed along filaments in some regular pattern. Interestingly, the $Z^2_1$ statistic shows that there is also a small maximum around 4~$h^{-1}$Mpc. This indicates that between two groups that are separated by 7~$h^{-1}$Mpc there is quite often another group.

Figure~\ref{fig:ray_cl} shows the $Z$-squared statistic for groups with different richness limits (at least one, two or three galaxies per group). The blue region shows the Monte Carlo simulation results for a Poisson sample. The $Z^2_1$ statistic for groups lies considerably below of that for the Poisson sample. This indicates that galaxy groups are distributed along filaments more regularly than in the Poisson case. This is expected, since galaxy groups cannot be located directly close to each other, there is some minimum distance between the groups that is also visible in the two-point correlation functions.  

In Fig.~\ref{fig:ray_cl} we see that the maxima around  4 and 7~$h^{-1}$Mpc that define the preferred scales for the galaxy distribution, are also visible in the group distributions. Interestingly, the 7~$h^{-1}$Mpc scale is the strongest for groups with at least three member galaxies. It shows that the regularity is stronger if the weakest groups and single galaxies are excluded.

\section{Concluding remarks}

Using the Bisous model (marked point process with interactions) we extracted the galaxy filaments from the SDSS spectroscopic galaxy survey. The diameter of the extracted filaments is roughly 1~$h^{-1}$Mpc and the catalogue of filament spines is built as described in \citet{Tempel:14a}. Using the galaxies and groups in filaments (with a distance from the filament axis less than 0.5~$h^{-1}$Mpc) we studied how the galaxies/groups are distributed along the filament axis. The main results of our study can be summarised as following.
\begin{itemize}
	\item The galaxy and group distributions along filaments show a regular pattern with a preferred scale around 7~$h^{-1}$Mpc. A weaker regularity is also visible at a scale of 4~$h^{-1}$Mpc. The regularity of the distribution of galaxies along filaments is a new result that might help to understand structure formation in the Universe.
	\item The pair correlation functions of galaxies and groups along filaments show that around each group, there is a region where the number density of galaxies/groups is smaller than on average.
	\item Galaxy groups in the Universe are more uniformly distributed along filaments than in the Poisson case.
\end{itemize}

The clustering pattern of galaxies and groups along filaments tells us that galaxy filaments are like pearl necklaces, where the pearls are galaxy groups that are distributed along the filaments in some regular pattern.

We suggest that the measured regularity of the galaxy distribution along filaments could be used as a cosmological probe to discriminate between various dark energy and dark matter cosmological models. Additionally, it can be used to probe environmental effects in the formation and evolution of galaxies. We plan to test these hypothesis in our following analysis using N-body simulations and deep redshift surveys like the Galaxy And Mass Assembly \citep[GAMA,][]{Driver:11} and the VIMOS Public Extragalactic Survey \citep[VIPERS,][]{Garilli:14}.

\begin{acknowledgements}
      We thank the Referee for the useful and constructive report that helped to improve the presentation of our results. This work was supported by institutional research funding IUT26-2, IUT40-2 and TK120 of the Estonian Ministry of Education and Research, and by the Estonian Research Council grant MJD272.  Funding for SDSS-III has been provided by the Alfred P. Sloan Foundation, the Participating Institutions, the National Science Foundation, and the U.S. Department of Energy Office of Science. The SDSS-III web site is http://www.sdss3.org/. SDSS-III is managed by the Astrophysical Research Consortium for the Participating Institutions of the SDSS-III Collaboration including the University of Arizona, the Brazilian Participation Group, Brookhaven National Laboratory, Carnegie Mellon University, University of Florida, the French Participation Group, the German Participation Group, Harvard University, the Instituto de Astrofisica de Canarias, the Michigan State/Notre Dame/JINA Participation Group, Johns Hopkins University, Lawrence Berkeley National Laboratory, Max Planck Institute for Astrophysics, Max Planck Institute for Extraterrestrial Physics, New Mexico State University, New York University, Ohio State University, Pennsylvania State University, University of Portsmouth, Princeton University, the Spanish Participation Group, University of Tokyo, University of Utah, Vanderbilt University, University of Virginia, University of Washington, and Yale University.
\end{acknowledgements}



\appendix

\section{Kernel density estimation}
\label{app:kernel}

\begin{figure}
   \centering
   \includegraphics[width=80mm]{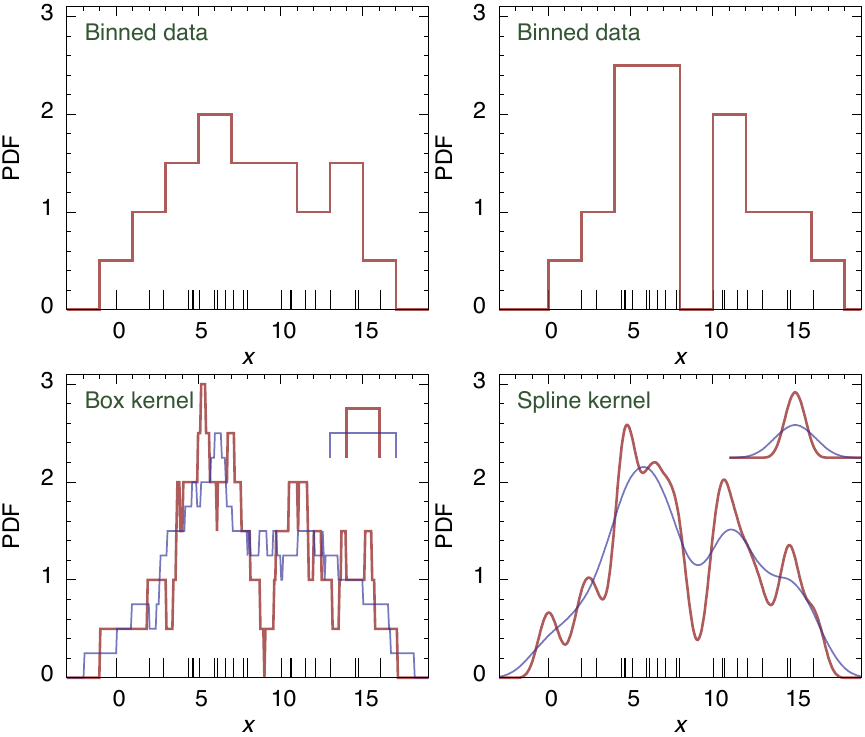}
   \caption{The upper row demonstrates probability density estimation using binning. In the left and right panels, the bin width is the same but the centre of the bin is shifted. The bottom row demonstrates density estimation using kernel smoothing. In the bottom left panel the simplest box kernel is used, in the bottom right panel the $B_3$ spline kernel is used. The kernel shape and size is shown  in the upper-right corner of the figures. The rug plot on the bottom axis shows the data points that were used for density estimation.}
   \label{fig:kernel}
\end{figure}

In the simplest case, the probability density can be estimated as the binned density histogram. However, this estimate depends both on the bin widths and the location of the bin edges. A better way is to use kernel smoothing \citep[e.g.][]{Silverman:86, Wand:95, Feigelson:12}, where the density is represented by a sum of kernels centred at the data points:
\begin{equation}
	f(x) = \frac{1}{h}\sum\limits_i K\left( \frac{x-x_i}{h} \right) .
\end{equation}
The kernels $K(x)$ are distributions ($K(x)>0$, $\int K(x)\mathrm{d}x=1$) of zero mean and of a typical width $h$. The width $h$ is an analogue of the bin width, but there are no bin edges to worry about.

To create our density estimation we use the popular $B_3$ spline function
\begin{equation}
	B_3(x) = \frac{ |x-2|^3 -4|x-1|^3 +6|x|^3 -4|x+1|^3 +|x+2|^3 }{12}.
\end{equation}
This kernel is well suited for estimating densities -- it is compact, differing from zero only in the interval $x\in [-2,2]$, and it conserves mass: $\sum_i B_3(x-i)=1$ for any $x$.

In many papers \citep[e.g.][]{Vio:94, Fadda:98, Ferdosi:11} it has been shown that kernel smoothing is the best and recommended choice for density estimation. It is more robust and reliable than other simpler methods and provides comparable (or better) results than methods with high computational costs (e.g. the  maximum likelihood technique).

To illustrate the use of kernel smoothing, in Fig.~\ref{fig:kernel} we show the estimated probability density function using either binning the data or kernel smoothing with two types of kernels (the box and $B_3$ spline kernels). The upper row in Fig.~\ref{fig:kernel} demonstrates density estimation using binning, where the bin widths are the same but the bin centres are slightly shifted. We see that the resulting density estimate may depend  strongly on the chosen bin locations.

The lower-left panel in Fig.~\ref{fig:kernel} demonstrates density estimation using the simplest box kernel. The kernel width is the same as box width for binned density estimates above. Even the simplest box kernel reveals the details in density distribution, however, the resulting distribution function is not smooth. To get a smooth density distribution, a smooth kernel should be used. The kernel density estimates using $B_3$ spline kernels are shown in the lower-right panel of Fig.~\ref{fig:kernel}. To show that the results of kernel smoothing are quite robust, we doubled the kernel width and compared the probability density estimates as shown in Fig.~\ref{fig:kernel}.

\section{Examples of the Rayleigh ($Z$-squared) statistic}
\label{app:z2}

\begin{figure}
   \centering
   \includegraphics[width=80mm]{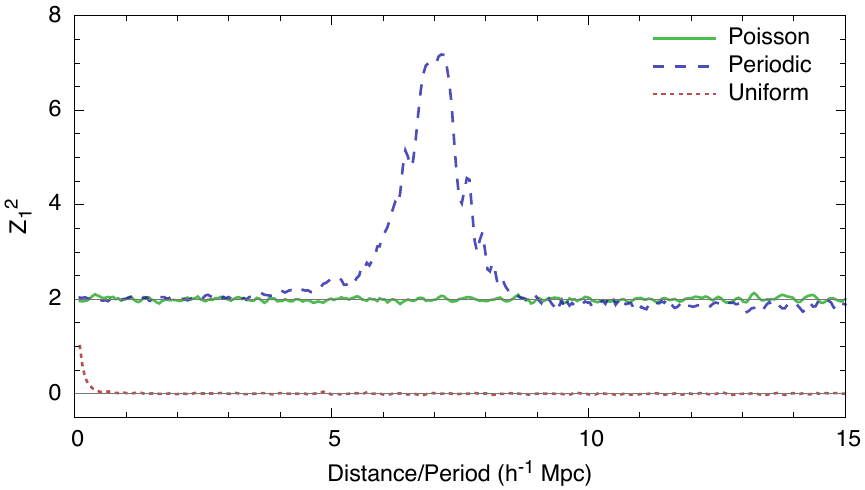}
   \caption{The $Z$-squared statistics for three cases. The green line shows the $Z$-squared statistic for  a Poisson sample. The blue line shows the statistic for a periodic signal, where the period for each datapoint is drawn from a Gaussian distribution centred at 7~$h^{-1}$Mpc with a standard deviation of 0.5~$h^{-1}$Mpc. The red line shows the statistic for data points  with an uniform point distribution -- see text for more information.}
   \label{fig:ray_test}
\end{figure}

To illustrate how the $Z$-squared statistic works, we generated three datasets and calculated the Rayleigh statistics for them. The results are shown in Fig.~\ref{fig:ray_test}. In the first case, we generated a Poisson distribution (green line). For a Poisson sample, the statistic gives an average value 2. In the second case, we added some periodicity to the sample (blue line). We generated point distributions with the same (zero) phase and with periods chosen from a Gaussian distribution centred at 7~$h^{-1}$Mpc with a standard deviation of 0.5~$h^{-1}$Mpc, and added these together. In Fig.~\ref{fig:ray_test} we see that the $Z$-squared statistic recovers the period well. In the third case, we generated a uniform distribution of points (red line). For that, we divided the test filament into $N$ (the number of points) equal regions and in each region we put one point, selected from a uniform distribution. Doing that we get points that are more homogeneously distributed along the filament and the $Z$-squared statistic gives the value zero. We can conclude that if the value of the statistic lies above 2, there is some periodicity in the data. Contrary, if it is below 2, it describes a more homogeneous distribution. Additionally, the peaks in the $Z$-squared statistic show that there is preferred periodicity in the data with the scale of the peak position.

It is known that for the uniform distribution of points, the  $Z$-squared statistic is distributed as the chi-square with two degrees of freedom \citep[see, e.g.][]{Wall:12}. So its expectation value is 2, as seen in Fig.~\ref{fig:ray_test}.

\end{document}